# Dialog interface for dynamic data models

**Mgr. Vojtěch Přehnal**

E-mail: vojtech.prehnal@gmail.com

Laboratory of Searching and Dialogue, Department of Information Technologies
Faculty of Informatics, Masaryk University, Brno, Czech Republic

**Abstract:** The vast majority of today's information systems do not provide ability to change the structure of the stored data (i.e. data model) at the run time ("on the fly"). When the user needs to change the data model, they have to contact the manufacturer of the system and wait for them to fulfill their requirements. This takes some time which may cause considerable financial loss for the company and may prevent user from laying up new claims, although they could bring them additional gains. In some cases it's necessary to stop the system for a while (during the data model adjustments), which may cause additional losses or may be completely impossible. Furthermore, the user has to pay for something they could make themselves easily, using a few clicks of mouse. The user is dependent on the supplier/manufacturer of the system, and if the contract is terminated, the possibility of any kind of maintenance is over in the fact. Last but not least, the most of workload in the information system development is concerned on simple, fully-automatable tasks, like updating data model, running ORM (object-relational mapping) tools, creating user interface for detailed record filtering and for individual data views, implementing common business logic for each table, such as CRUD operations (Create, Read, Update, Delete), data serialization and transport over the network.

In this paper, the new information system development methodology will be proposed. This methodology will enable the whole data model to be built and adjusted at the run time, without rebuilding the application. This will make the user much more powerful and independent on the manufacturer of the system. It will also cut the price and shorten the development time of the information systems dramatically, because common business logic will not have to be implemented for each individual table and the major part of the user interface will be generated automatically.

## 1. Introduction

### 1.1 Terminology

In this paper, the following terminology will be used:

- **Item**: an ordered multi-set of related data values. The rank of each data value is referred to as an **ordinal**
- **Field**: a multi-set of data values with the same ordinal and some common attributes (type, size, nullability, identity, …)
- **Table**: a multi-set of data values organized using a model of **items** (rows, records) and **fields** (columns)
- **Data model:** a set of tables and their relations
- **Dynamic data model:** data model with time-varying data structure, fully editable at the run-time
- **Dialog interface:** communication protocol between client and server
- **Meta-data:** supplementary information defining the structure and the attributes of the raw data
- **CRUD operations:** standard database operations with the items: Create (insert), Read (select), Update and Delete.

## 1.2 Three-tier architecture

The main purpose of today's information systems is to provide access to the enterprise data and to support various business processes. This is where the three-tier client-server architecture comes into the play.

Three-tier architecture is well-established and recommended architecture and software design pattern suitable for the vast majority of today's information systems and business scenarios. **Presentation tier** (typically a web client or a desktop application running on client machine) provides user interface for entering inputs and displaying outputs. **Application (or logic) tier** (typically web services and/or applications) performs detailed data processing on the server supporting desired business logic. **Data tier** consists of one or more databases running on one or more database servers. Here all the information is stored and retrieved.

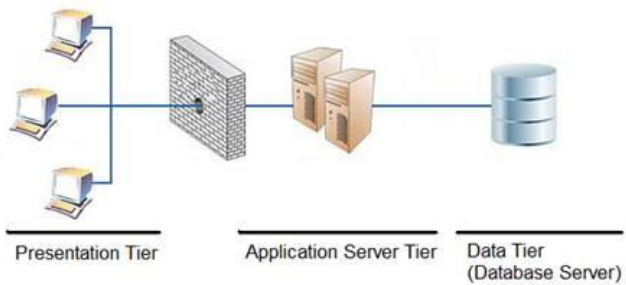

These three tiers are deeply related to each other. When the data model is adjusted, the business logic needs to be adjusted too in most cases, as well as the user interface. Hence, the changes in the data tier imply additional changes in application and presentation tier. In other words: **three-tier architecture means three-tier development** resulting in spending extra time and money for each single tier and data model adjustment.

## 1.3 Compile-time development automatization

There are some tools for simplifying the most common development routines regarding changes in the data model in the application and presentation tier. For example, ORM tools (object-relational mapping) such as Microsoft ADO.NET [2] or Microsoft Linq2Sql [3] together with Microsoft WCF RIA Services [4] provide both automatic class model and application logic generation, which may simplify application tier development substantially. Several advanced development tools like Microsoft ASP.NET Dynamic Data [5] or Microsoft WebMatrix [6] actually provide automatic generation of all the application source code directly from the data model.

However, the output of all these tools is always some kind of source code and thus they provide only compile-time development automatization and they always **require rebuilding the application** in the end. They are simply not able to make the running application know about performed changes in the data model. Furthermore, these tools introduce a number of other compelling limitations while adjusting the data model. For example, in the most of these tools the whole relevant source code is regenerated after the changes in the data model, which may result in loss of specialized or customized business logic supplied manually by the programmer.

## 1.4 Run-time data model exploration

The goal of this paper is to propose a new software development methodology that enables the running applications to access the data without any explicit knowledge of their model. To do this, it's necessary to acquire the information about the data model automatically at the run-time, prior to accessing the data.

However, retrieving the information about the whole data model at once is a very expensive operation because the data model may contain hundreds or thousands of data tables. Moreover, the acquired information about the data model may suffer from deterioration, because the data model is time-varying, thorough we need this information utmost up-to-date in order to perform desired data transactions successfully. The solution is to acquire only the appropriate subset of the data model prior to the desired data operations within a single transaction. This way, the accuracy of the acquired information of the data model is guaranteed and the amount of this information is limited to the constant size.

The communication between client and server pass over the conventional request-response protocol. The client sends a **request** for the desired subset of the database and the server sends the specified **data** together with their **model** in a **response**. This attached **meta-data** contains the information about the structure and all the properties of the data retrieved from the server. This way the client is able to display data properly, to perform basic client-side data validation logic and to request the server for the desired operations.

These requests and responses exchanged between the client and the server represent the sequential series of messages arranged successively in time forming a kind of dialog interface between the client and the server. **Using this dialog interface, the data model can be explored interactively** rather than being acquired as a whole and downloaded once at time. In this paper, the suitable meta-data model will be discussed and designed in order to provide all the necessary meta-information to client.

## 2. Methodology

### 2.1 Objectives

The main objective of this methodology is to propose a new approach to information system (or, generally, to data-driven application) development enabling editing data model at the run-time. This methodology should substitute conventional methodologies, practices, tools and design patterns based on three-tier model preserving all their functionality and adaptability.

To achieve these goals, a communication foundation will be designed providing **asynchronous message exchange** between client and server. A **dialog interface** consisting of the data model of these messages and their sequential arrangement in time will be defined and a suitable implementation of related application and presentation tier will be recommended. For hand-on examples of platform-independent data model retrieval using SQL, the standardized view **INFORMATION_SCHEMA** (defined by SQL-92 standard [7]) will be utilized [8].

### 2.2 Reading from database

The crucial functionality of each information system is to retrieve data from the database and to present these data in a way the user can simply use and understand. The most used way of presenting data is so called **master-details** view. User can browse the items of a given table and edit their attributes.

#### 2.2.1 Reading list of tables

In the dynamic data model, the list of tables may change in time. Therefore the first functionality the application tier has to provide is retrieving the current list of tables in the database. This can be performed using standardized **INFORMATION_SCHEMA** view like this:

**SQL:**

```sql
select [information_schema].[tables].[table_name]
from   [information_schema].[tables]
where  [information_schema].[tables].[table_type] = 'base table';
```

**Communication diagram:**　　　　　　　　　　　　**Class diagram:**

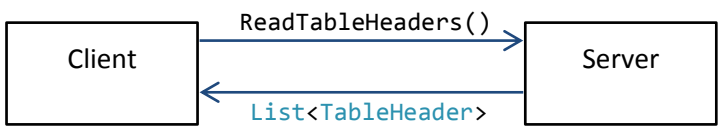　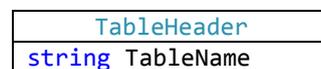

**Class description:**

- **TableName** contains the name of the table in the database.

**Presentation tier:**

The client should present the retrieved list in the form of vertical or horizontal menu, thus the navigation user interface may be automatically generated herewith. The user chooses the desired table from the menu and the appropriate table with all the meta-information (such as list of its fields and relations) should be retrieved and presented on demand. This way, the user may browse all the tables in the database and explore the whole data model interactively.

## 2.2.2 Reading list of fields (columns) in table

The most important meta-information about the specified table is the current list of its fields together with their attributes. For this purpose, **INFORMATION_SCHEMA** view may be utilized this way:

**SQL:**

```sql
use insight;
select      [information_schema].[columns].[column_name],
            [information_schema].[columns].[table_name],
            [information_schema].[columns].[data_type],
            [information_schema].[columns].[is_nullable],
            [information_schema].[columns].[character_maximum_length],
            [information_schema].[table_constraints].[constraint_type],
            [information_schema].[key_column_usage].[table_name] as [pk_table_name],
            [information_schema].[key_column_usage].[column_name] as [pk_column_name]
from        [information_schema].[columns]
left join   [information_schema].[key_column_usage]
            on [information_schema].[columns].[column_name] =
               [information_schema].[key_column_usage].[column_name]
            and [information_schema].[columns].[table_name] =
               [information_schema].[key_column_usage].[table_name]
left join   [information_schema].[table_constraints]
            on [information_schema].[key_column_usage].[constraint_name] =
               [information_schema].[table_constraints].[constraint_name]
where       [information_schema].[columns].[table_name] = @table_name
order by    [information_schema].[columns].[ordinal_position]
```

**Communication diagram:**

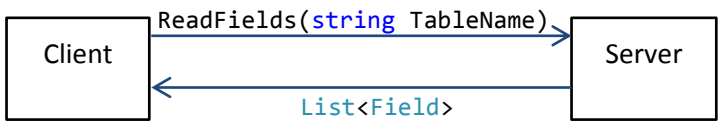

**Class diagram:**

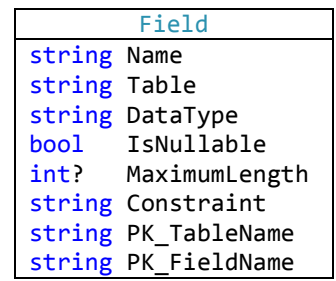

**Class description:**

- **Name** - the name of the field
- **Table** - the name of the parent table
- **DataType** - the data type in textual form ("int", "varchar", "bit" …)
- **MaximumLength** - the maximum character length of text fields (for other fields this attribute is null and for unlimited lengths is equal to -1)
- **IsNullable** - holds true if the field may contain null values
- **Constraint** - type of constraint on the field ("PRIMARY KEY", "FOREIGN KEY", "CHECK", "UNIQUE")
- **PK_TableName** - name of referenced table with primary key (only for fields with "FOREIGN KEY" constraint)
- **PK_FieldName** - name of referenced field with primary key (only for fields with "FOREIGN KEY" constraint)

**Presentation tier:**

The client should display a table filled with columns of the same name as the retrieved fields. After the user opens the given item (row), the details view form should present individual data values together with appropriate field names and enable editing and submitting data back to the server. Basic validation logic should be performed by the client using the knowledge about the data types and other meta-information provided in this list.

### 2.2.3 Reading list of relations

It is also useful to know the current list of tables referencing the specified table. This list may be presented as an auto-generated submenu of the open item enabling the user to navigate to matching items in the referencing tables. This can be achieved by this SQL query on **INFORMATION_SCHEMA** view:

**SQL:**

```sql
use insight;
select      [foreign_keys].[table_name] as [fk_table_name],
            [foreign_keys].[column_name] as [fk_column_name],
            [private_keys].[column_name] as [pk_column_name]
from        [information_schema].[key_column_usage]  as [foreign_keys]
inner join  [information_schema].[table_constraints] as [foreign_constraints]
            on  [foreign_keys].[constraint_name]
                = [foreign_constraints].[constraint_name]
            and [foreign_constraints].[constraint_type] = 'FOREIGN KEY'
inner join  [information_schema].[referential_constraints]
            on  [foreign_constraints].[constraint_name]
                = [information_schema].[referential_constraints].[constraint_name]
inner join  [information_schema].[table_constraints] as [private_constraints]
            on  [information_schema].[referential_constraints].[unique_constraint_name]
                = [private_constraints].[constraint_name]
inner join  [information_schema].[key_column_usage]  as [private_keys]
            on  [private_constraints].[constraint_name]
                = [private_keys].[constraint_name]
            and [foreign_keys].[ordinal_position] = [private_keys].[ordinal_position]
where       [private_constraints].[table_name] = @table_name
order by    [fk_table_name]
```

**Communication diagram:**

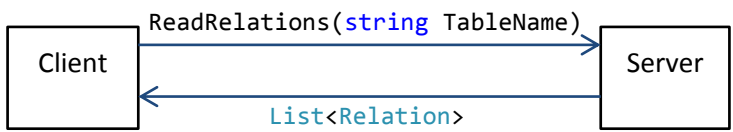

**Class diagram:**

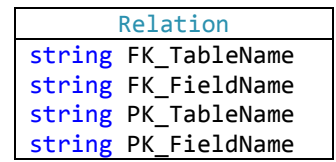

**Class description:**

- **FK_TableName** - the name of foreign-key (referencing) table
- **FK_FieldName** - the name of foreign-key (referencing) field
- **PK_TableName** - the name of private-key (referenced) table (always the same as **TableName** parameter)
- **PK_FieldName** - the name of private-key (referenced) field

**Presentation tier:**

While displaying the details view form of the open item, the client should display a menu consisting of referencing tables navigating to matching items in these tables. If there are more referencing fields within one referencing table, for each referencing field there should be appropriate sub-item in the menu.

### 2.2.4 Reading items (rows) from table

Now, with the knowledge of table fields and relations, it is possible to perform basic reading from the table, i.e. to read all the items in the table like this:

**SQL:**

```sql
select    *
from      @table_name
where     -- here comes FilterExpression
order by  -- here comes OrderExpression
```

| Communication diagram: | Class diagram: |
|---|---|
| 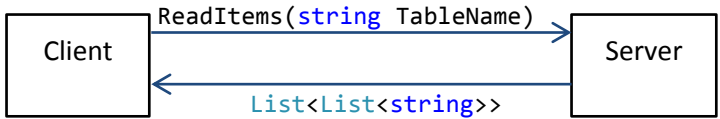 | No need for special class, all data values should be converted to string and returned as two-dimensional array of strings in order to prevent problems with the serialization over the network. |

For detailed item selection (such as filtering, paging or ordering), the header of ReadItems method should be extended this way:

`ReadItems(string TableName, int Skip, int Take, string OrderExpression, string FilterExpression)`

where Skip and Take are paging parameters (the index of the first item and the number of items to retrieve), OrderExpression is the content of SQL **order by** clause and FilterExpression is the content of SQL **where** clause. These parameters have to be included in appropriate clauses in underlying SQL query.

## 2.2.5 Reading total count of items passing given criteria

If the paging is performed, it's necessary to get the total number of the items in the table although not all the items are retrieved. For this purpose, a simple query should be used:

**SQL:**

```sql
select   count(*)
from     @table_name
where    -- here comes FilterExpression
```

**Communication diagram:**

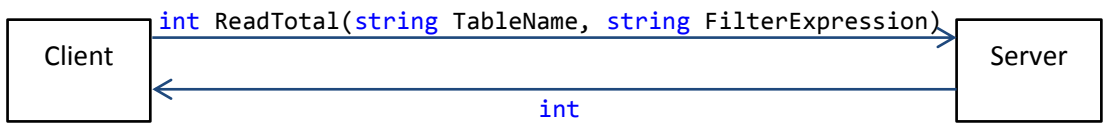

**Presentation tier:**

The client should display a total number of items passing specified filter at the suitable position (e.g. above or under the table).

## 2.2.5 Communication protocol optimization

As far as here, these methods for data and meta-data retrieval were described:

```
ReadTableHeaders()
ReadFields(string TableName)
ReadRelations(string TableName)
ReadItems(string TableName, int Skip, int Take, string OrderExpression, string FilterExpression)
ReadTotal(string TableName, string FilterExpression)
```

Although in theory it is possible for the client to make a request for each of these operations, performing all these operations at once within a **single transaction** on the server results in much more efficient processing, ensures consistency of retrieved data and meta-data and saves a couple of redundant request-response round-trips over the network. At the start, the client retrieves the list of tables (which is basically invariable) from the server and then sends individual requests to the server for each table requested by the user.

Prior to processing mentioned operations, a new connection to the database should be established and a new database transaction should be started. This connection and transaction has to be specified as additional parameters of each individual operation. These attributes were omitted in the diagram below for better readability.

**Dialog interface for single-request single-transaction table retrieval:**

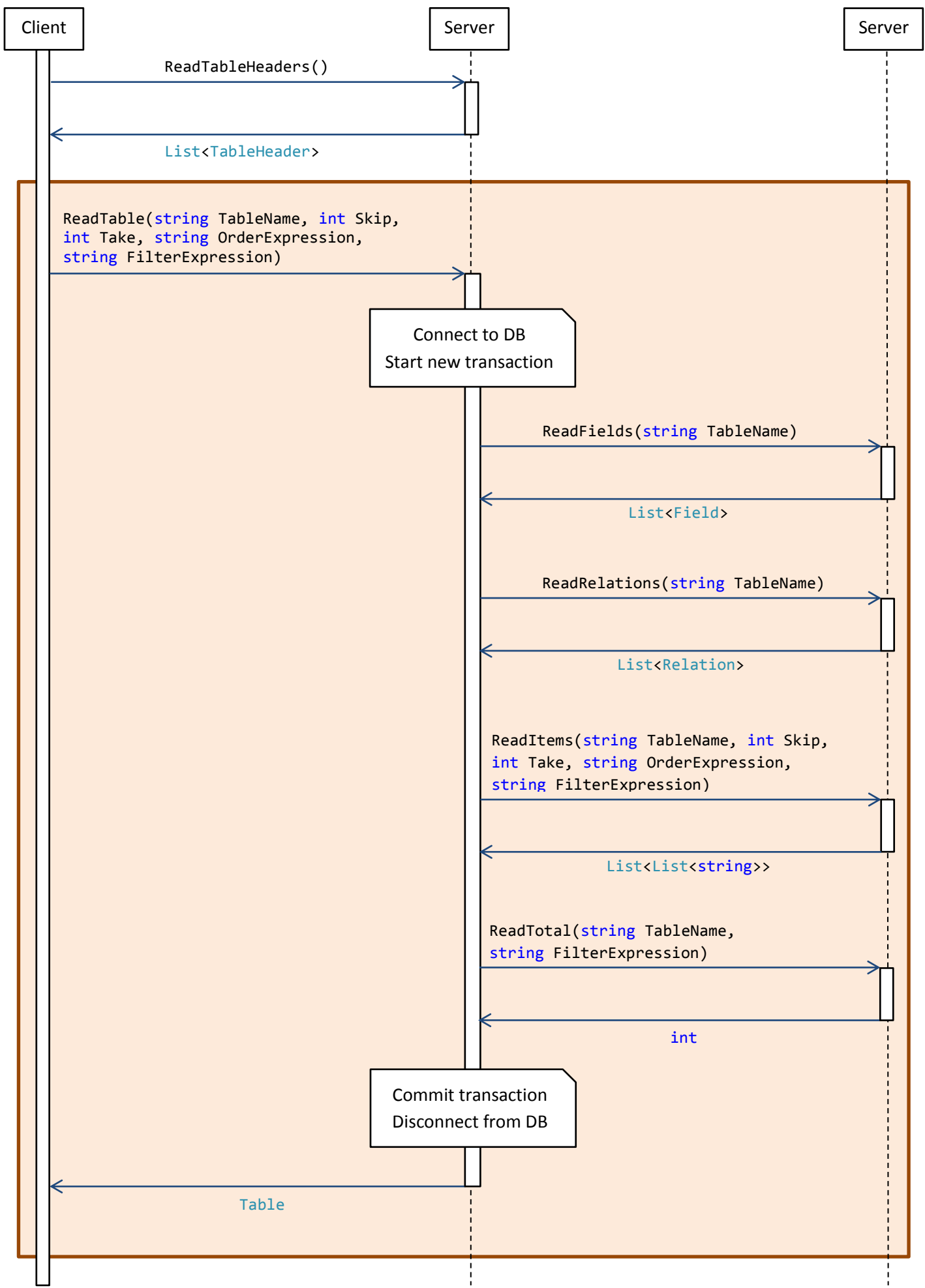

**Class diagram:**

```
            Table
List<Field> Fields
List<Relation> Relations
List<List<string>> Items
int Total
```

**Class description:**

- **Fields** - the list of fields
- **Relations** - the list of relations
- **Items** - the list of items
- **Total** - total number of items passing specified filter expression

**Presentation tier:**

After receiving the complete data about the desired table, the client should display table filled with columns and populated with desired data (items). The total number of items should be displayed near the table. After opening a given item from the table, the details view form should be displayed enabling data editing and basic data validation.

# 3. Conclusion and future work

In this paper, the new methodology of information system development was introduced. The key point of this methodology is **retrieving data model at the runtime** instead of building a fixed data model at compile-time. The main advantage of this approach is ability to **edit data model on the fly** with **no need of rebuilding** or restarting the running application. This brings a whole new level of flexibility and customization to the world of information.

The basic data operations for dynamic data model were suggested. It was shown, how to retrieve the most important meta-data from the database using the standardized **INFORMATION_SCHEMA** view. The data model of request and response messages was designed and the dialog interface between client and server for efficient data retrieval was presented.

The proposed methods, messages, data models and communication interface are **easily extensible** to comprehend any special requirement and some important extensions are to be presented in oncoming articles (see Future work chapter). **In the fact there's no reason why all the information systems and other data-driven application should not work like this. Therefore we propose this method as a new standard for information system development.**

# 4. Future work

Some useful extensions of this methodology are planned to be introduced in oncoming articles. These extensions include:

- Auto-join and display-fields (very useful techniques for displaying meaningful data values instead of key values in foreign-key fields)
- Submitting data changes to database
- Specialized business logic implementation
- Utilizing Microsoft Ribbon for user interface design
- Data access management
- User rights management
- Data model and user interface localization
- Automatic SQL-injection protection
- Automatic data log creation

# 5. Acknowledgement

The author would like to thank to the supervisor for his helpful discussions, hints and advices.

**Mgr. Vojtěch Přehnal** (author) is a Ph.D. candidate at Faculty of Informatics, Masaryk University, Brno, Czech Republic. He has over five years of hands-on experience in information system development. He is also invited reviewer for… His research interests include information system development, management and enterprise costs optimization.

**Doc. RNDr. Ivan Kopeček, CSc.** (supervisor) is an associate professor at Department of Information Technologies, Faculty of Informatics, Masaryk University, Brno, Czech Republic. He is also a leader of Laboratory of Speech and Dialogue and a member of Branch Council of the faculty.